\documentclass[aps,apl,twocolumn,superscriptaddress]{revtex4}
\usepackage{graphicx}
\usepackage{epstopdf}
\usepackage{comment}
\usepackage{amsmath,amssymb}
\begin{document}

\title{Cavity magnomechanics with surface acoustic waves}

\author{D. Hatanaka}

\email{daiki.hatanaka.hz@hco.ntt.co.jp}

\author{M. Asano}

\author{H. Okamoto}

\author{Y. Kunihashi}

\author{H. Sanada}

\author{H. Yamaguchi}

\affiliation{NTT Basic Research Laboratories, NTT Corporation, Atsugi-shi, Kanagawa 243-0198, Japan}



\maketitle

\hspace*{0.5cm}\textbf{Magnons, namely spin waves, are collective spin excitations in ferromagnets \cite{kittel}, and their control through coupling with other excitations is a key technology for future hybrid spintronic devices \cite{tabuchi_PRL2014,tang_magnon,bai_PRL2015,tabuchi_Science2016,hisatomi_PRB2016,hou_PRL2019,li_PRL2019,nakamura_science2020}. Although strong coupling has been demonstrated with microwave photonic structures, an alternative approach permitting high density integration and minimized electromagnetic crosstalk is required. Here we report a planar cavity magnomechanical system, where the cavity of surface acoustic waves enhances the spatial and spectral power density to thus implement magnon-phonon coupling at room temperature. Excitation of spin-wave resonance involves significant acoustic power absorption, whereas the collective spin motion reversely exerts a back-action force on the cavity dynamics. The cavity frequency and quality-factor are significantly modified by the back-action effect, and the resultant cooperativity exceeds unity, suggesting coherent interaction between magnons and phonons. The demonstration of a chip-scale magnomechanical system paves the way to the development of novel spin-acoustic technologies for classical and quantum applications.}\\
\hspace*{1.0em}Acoustic phonons allow the interconnection between different physical systems and has attracted attention, especially in the fields of cavity optomechanics \cite{aspel_revmodphys,cohadon_PRL1999,kippenberg_optomech_rev,connell_qgs,chan_qgs,teufel_qgs} and circuit quantum acoustodynamics (c-QAD) \cite{soykal_qad,delsing_nphys,delsing_science,schuetz_prx,manenti_qae,chu_qae,noguchi_saw,cleland_saw,safavi_pnc2}. In contrast to microwaves, acoustic waves have a short micrometer-scale wavelength at gigahertz frequencies and no radiation loss into free space. In addition, hybrid devices are highly tunable and thus enable coherent signal manipulation in magnonic systems. Although the ability to drive spin-wave resonance by traveling surface acoustic waves (SAWs) has been extensively investigated \cite{weiler_adfmr,dreher_adfmr,thevenard_adfmr1,labanowski_adfmr,kobayashi_src,onose_PRB2019,santos_magnomech2020,otani_APL2020}, the features of magnon-phonon coupling, i.e. the coherent magnon excitation by acoustic means and the back-action effect on acoustic resonance, have yet to be fully explored due to their weak mutual interaction.\\
\begin{figure*}[t]
	\begin{center}
		\vspace{-0.2cm}\hspace{-0.0cm}
		\includegraphics[scale=1]{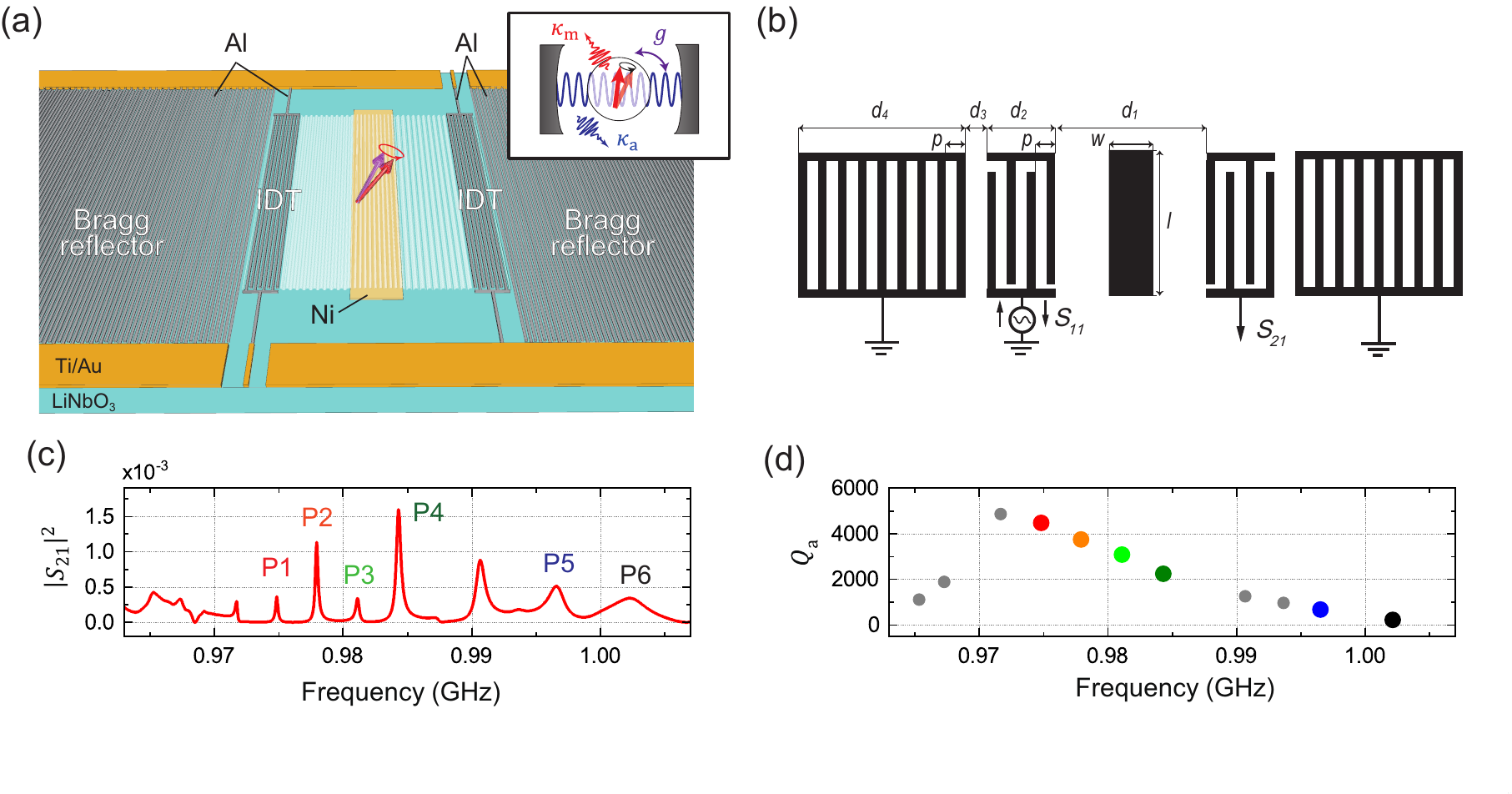}
		\vspace{-1.0cm}
		\caption{Schematic of cavity magnomechanical system based on a SAW resonator on a LiNbO$_{3}$ substrate. Acoustic Bragg reflectors composed of a one-dimensional periodic structure made of aluminum thin films are formed outside of both IDTs, which enables acoustic waves to be confined via the bandgap effect. A rectangular Ni film is deposited between the IDTs. A static magnetic field ($\mu_{0} H_{\rm ex}$) is applied in plane of the film and it aligns the magnetic moments. The inset is a schematic diagram of a cavity magnomechanical model where a magnonic cavity, i.e., spin-wave resonance, with angular frequency $\omega_{\rm m}$ and damping $\kappa_{\rm m}$ is interacted via coupling $g$ to an acoustic cavity with damping $\kappa_{\rm a}$. \textbf{(b)} The device geometry is shown, where $d_{1}$ = 557 $\mu$m, $d_{2}$ = 77 $\mu$m, $d_{3}$ = 2.9 $\mu$m, $d_{4}$ = 1137 $\mu$m, $p$ = 2.0 $\mu$m, $w$ = 120 $\mu$m and $l$ = 750 $\mu$m. \textbf{(c)} and \textbf{(d)} Frequency dependence of the SAW transmission ($|S_{21}|^{2}$) and quality-factor ($Q_{\rm a}$) are shown respectively. FP resonances denoted by solid color circles in (d) are labeled P1 (red), P2 (orange), P3 (light green), P4 (green), P5 (blue), and P6 (black).}
		\label{fig 1}
		\vspace{-0cm}
	\end{center}
\end{figure*}
\begin{figure}[t]
	\begin{center}
		\vspace{-0.2cm}\hspace{-0.0cm}
		\includegraphics[scale=0.95]{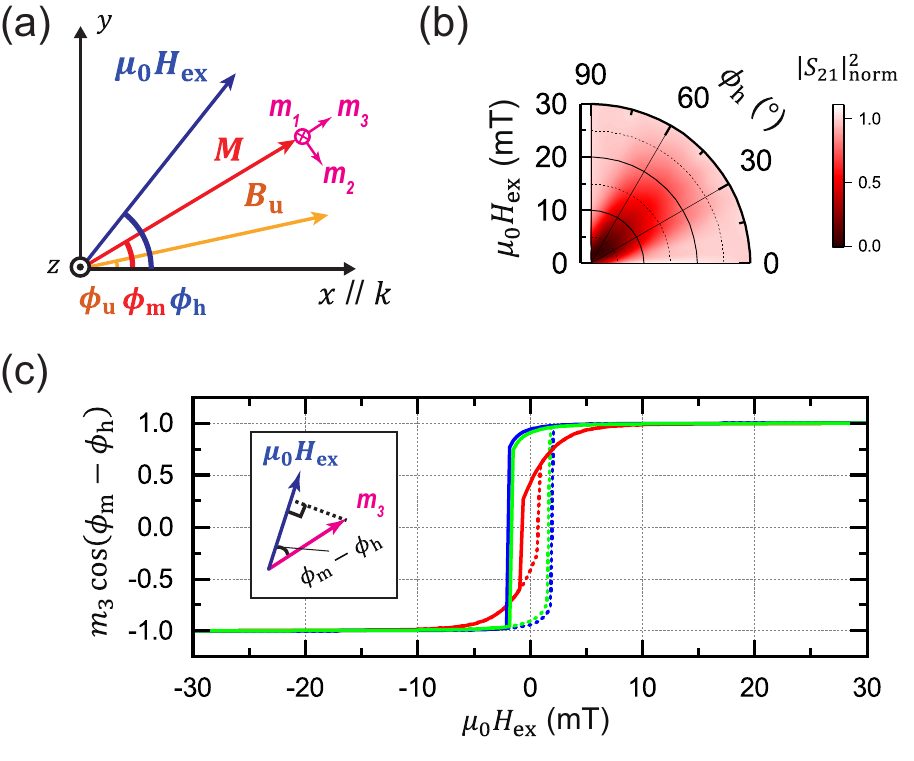}
		\vspace{-0.0cm}
		\caption{\textbf{(a)} Coordinates of external magnetic field ($\mu_{0} \mbox{\boldmath $H_{\rm ex}$}$), magnetization ($\mbox{\boldmath $M$}$), and uniaxial anisotropic field ($\mbox{\boldmath $B_{\rm u}$}$) with angle $\phi_{\rm h}$, $\phi_{\rm m}$ and $\phi_{\rm u}$ with respect to the $x$-axis, which is parallel to the SAW wavevector ($k$). Alternative coordinate system with $\mbox{\boldmath $m$}$ = ($m_{1}, m_{2}, m_{3}$), where $m_{3}$ is aligned to the equilibrium magnetization orientation. \textbf{(b)} Normalized acoustic transmission $|S_{21}|_{\rm norm}^{2}\equiv|S_{21}(\omega)|^{2}/|S_{21}(\omega)|_{\rm ref}^{2}$	in P1 resonance as function of an external field $\mu_{0} H_{\rm ex}$ measured at frequency $\omega/2\pi$ = 0.9748 GHz as the field angle is changed from $\phi_{\rm h}$ = 0$^\circ$ and 90$^\circ$. Here, $|S_{21}(\omega)|_{\rm ref}$ is the value at reference field $\mu_{0} H_{\rm ex}^{\rm ref}=$ 30 mT, which is chosen because of off-resonance condition of the magnon-phonon coupling. The acoustic power suppression due to spin-wave resonance excitation is observed around $\mu_{0} H_{\rm ex}$ = 2.1 mT at $\phi_{\rm h}$ = 45$^\circ$. \textbf{(c)} Static magnetization dynamics as function of the field $\mu_{0} H_{\rm ex}$ at $\phi_{\rm h}$ = 90$^\circ$ (red), 45$^\circ$ (blue), and 0$^\circ$ (green) calculated with the Stoner-Wohlfarth model, where the projection of $m$ to the field direction ($\cos(\phi_{\rm m}-\phi_{\rm h})$) as denoted in the left inset, is plotted on the $y$-axis. The results obtained during backward and forward sweeps are shown as solid and dotted lines respectively.}
		\label{fig 1}
		\vspace{-0.5cm}
	\end{center}
\end{figure}%
\begin{figure}[t]
\begin{center}
	\vspace{-0.0cm}\hspace{-0.0cm}
	\includegraphics[scale=1]{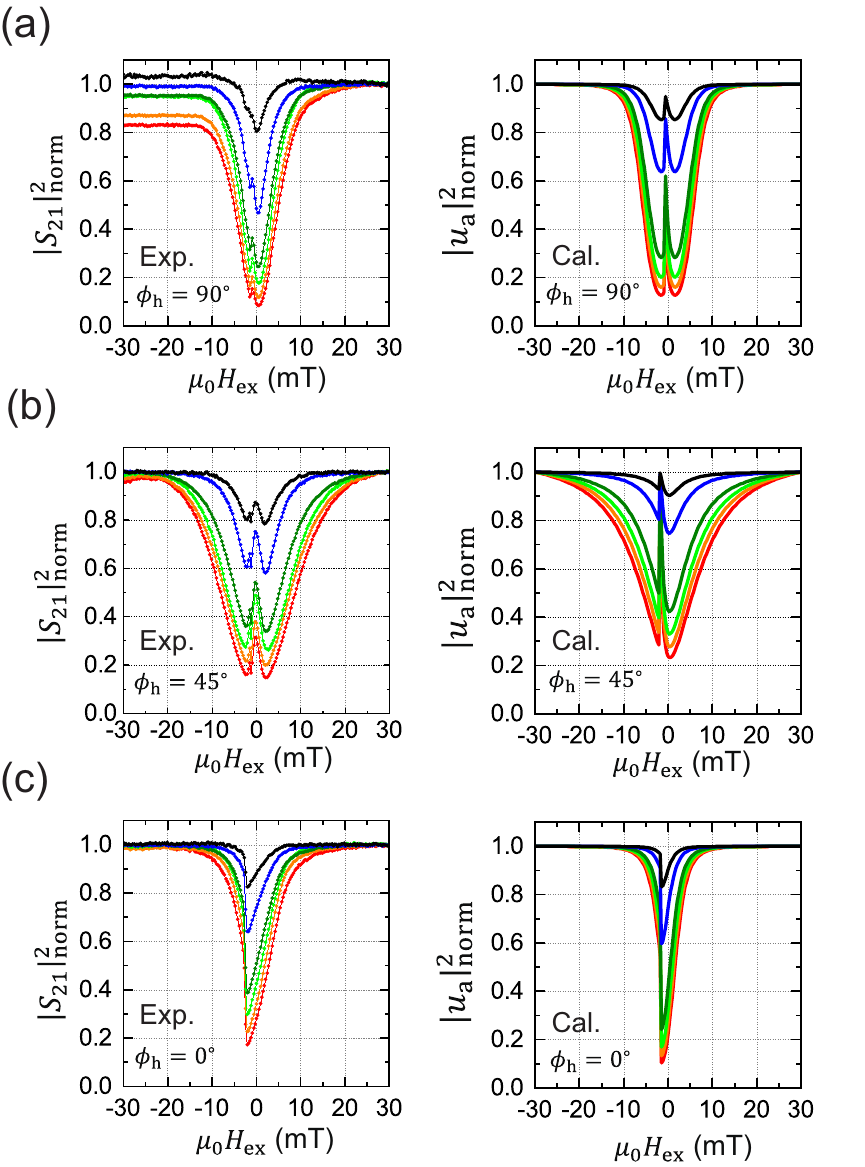}
	\vspace{-0cm}
	\caption{\textbf{(a), (b)} and \textbf{(c)} Field $\mu_{0} H_{\rm ex}$ response of the acoustic transmission measured in P1--P6 resonances with various $Q_{\rm a}$ in $\phi_{\rm h}$ = 90$^\circ$ (a), 45$^\circ$ (b) and 0$^\circ$ (c), obtained from the experiment (left) and the calculation (right). The results on the FP resonance with $Q_{a}$ = 4,500 (P1, red), 3,800 (P2, orange), 3,200 (P3, light green), 2,300 (P4, green), 700 (P5, blue), and 200 (P6, black) are plotted, whose colors correspond to those in Fig. 1(d). The normalized acoustic vibration is calculated by $|u_{\rm a}|_{\rm norm}^{2}=|u_{\rm a}|^{2}/|u_{\rm a}|_{\rm ref}^{2}$, where $|u_{\rm a}|_{\rm ref}$ is the value at $\mu_{\rm 0} H_{\rm ex}$ = 30 mT. Dual dips at $\phi_{\rm h}$ = 90$^\circ$ and 45$^\circ$ and a single dip at $\phi_{\rm h}$ = 0$^\circ$ are generated as a consequence of the acoustic excitation of spin-wave resonance.}
	\label{fig 1}
	\vspace{-0cm}
\end{center}
\end{figure}
\begin{figure*}[t]
	\begin{center}
		\vspace{-0cm}\hspace{-2.0cm}
		\includegraphics[scale=1.0]{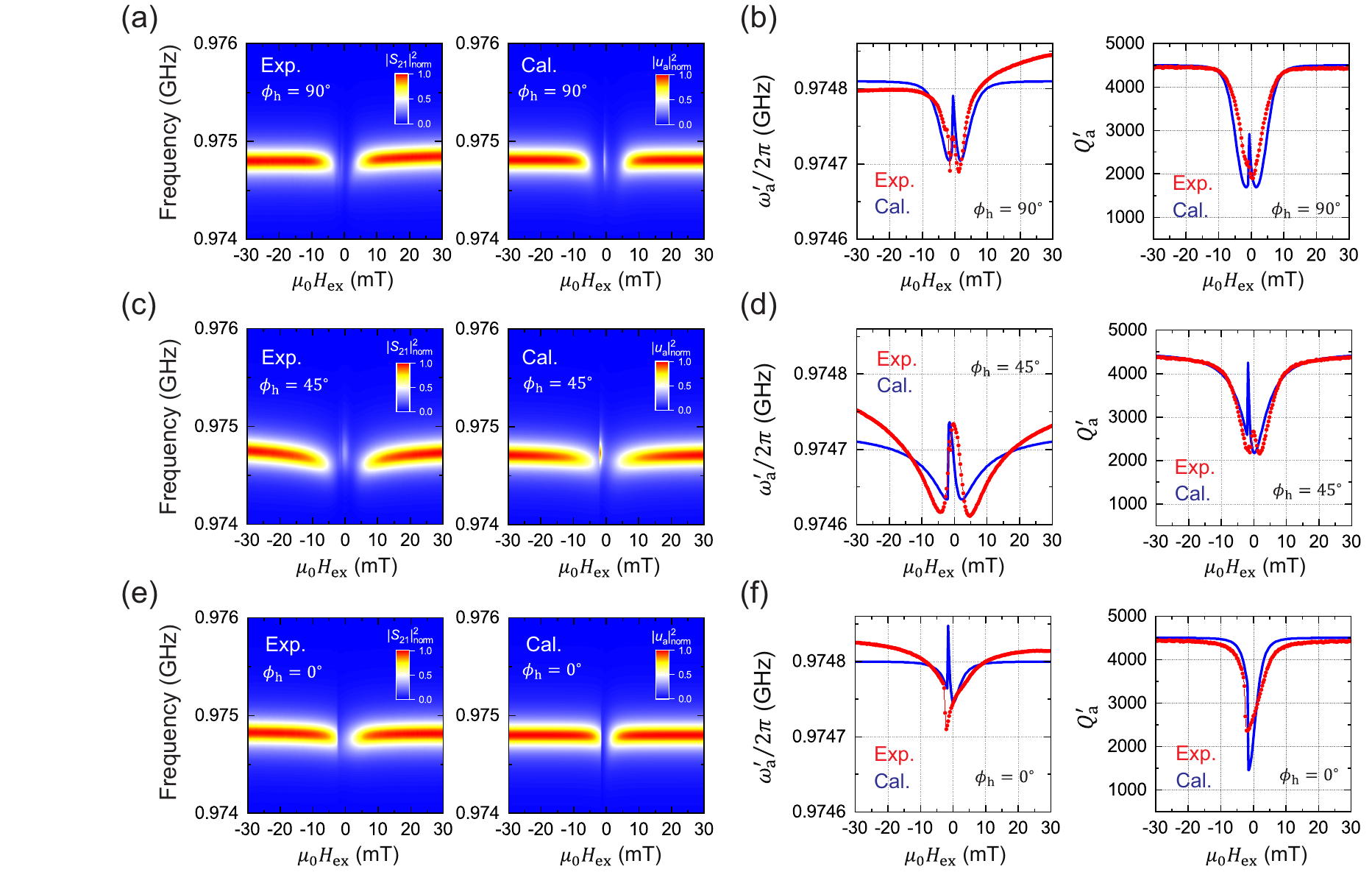}
		\vspace{-0cm}
		\caption{\textbf{(a), (c)} and \textbf{(e)} Left: Spectral responses of the acoustic resonance P1 at $\phi_{\rm h}$ = 90$^\circ$, 45$^\circ$ and 0$^\circ$, respectively, as the field was swept from $\mu_{0} H_{\rm ex}$ = 30 mT to -30 mT. The normalized magnitude $|S_{21}|_{\rm norm}^{2}$ is obtained from the value divided by that at $\mu_{0} H_{\rm ex}^{\rm ref}$ = 30 mT and at each resonance frequency. The different resonance frequency between $\phi_{\rm h}$ = 90$^\circ$, 45$^\circ$ and 0$^\circ$ is because the effect of a static stress induced by the magnetostriction is variant with changing field direction. Right: Simulated spectral response of the acoustic magnitude $|u_{\rm a}|_{\rm norm}^{2}$ at the corresponding $\phi_{\rm h}$. The resonance frequency used in the calculation was set to $\omega_{\rm a}/2\pi$ = 0.97481, 0.97473, and 0.97480 GHz respectively. \textbf{(b), (d)} and \textbf{(f)} Resonance frequency $\omega_{\rm a}^{\prime}/2\pi$ (left) and acoustic quality-factor $Q_{\rm a}^{\prime}$ (right) as function of field $\mu_{0} H_{\rm ex}$ at $\phi_{\rm h}$ = 90$^\circ$, 45$^\circ$ and 0$^\circ$, respectively. The experimental values obtained by fitting a Lorentzian curve to the measured spectra in (a)--(c) are shown by red solid-circles. The calculation data obtained from Eqss (2) and (3) described in the text and Methods is shown by the blue solid line.}
		\label{fig 1}
		\vspace{-0cm}
	\end{center}
\end{figure*}
\hspace*{0.5em}We have developed a planar cavity magnomehanical system on a LiNbO$_{3}$ substrate as shown in Figs. 1(a) and 1(b), which is piezoelectrically excited through an inter digital transducer (IDT), and the vibrations are resonantly enhanced by Bragg reflectors, forming a cavity \cite{tang_sawpnc,loncar_sawpnc}. The magnetization of a nickel (Ni) film is driven by the acoustic waves via a magnetostriction, generating collective spin resonance. This excitation dynamics is investigated by electrically measuring SAW transmission ($S_{21}$) at room temperature and in vacuum. The details on the device and measurement configuration are discussed in Methods.\\
\hspace*{1.0em}First, the spectral response of the acoustic cavity is investigated by exciting SAWs through an IDT and measuring them from another IDT as shown in Fig. 1(c). Multiple peaks appear in the frequency range from 0.963 to 1.007 GHz, where they are equally separated by $\Delta f$ = 3.0 MHz. The SAW velocity of LiNbO$_{3}$ is estimated to be $v_{\rm SAW}=\omega_{\rm a}/k$ = 3900 m/s from measured resonance angular frequency $\omega_{\rm a}$ and wavenumber $k$, which is consistent with a previous report \cite{LiNbO_vsaw}. Then, the cavity length is $L_{\rm c}=v_{\rm SAW}/(2\Delta f)$ = 650 $\mu$m, which almost corresponds to $d_{1}+2d_{2}$, indicating that the observed peaks are Fabry-Perot (FP) resonances by two Bragg reflectors. Figure 1(d) shows the quality-factor of each resonance ($Q_{\rm a}$) as a function of frequency, where it increases up to 5,000 in the frequency range between 0.970 and 0.975 GHz. The enhancement in $Q_{\rm a}$ results from the bandgap formed in the reflectors, which can be predicted by the finite-element method as described in the Supplemental Information (Section I). Thus, incorporating the acoustic reflectors into the system suppresses energy dissipation and provides high-quality acoustic cavity.\\
\hspace*{1.0em}Acoustic excitation of the spin-wave resonance is then demonstrated by measuring acoustic transmission ($S_{21}$) at 0.9748 GHz in the FP resonance labeled P1 [see Fig. 1(c)] while sweeping both the strength [$\mu_{0}H_{\rm ex}$] and directed angle [$\phi_{\rm h}$ defined in Fig.2(a)] of the external in-plane magnetic field. Figure 2(b) shows the transmission magnitude $|S_{21}|^{2}$ normalized by that under the off-resonance condition at $\mu_{0} H_{\rm ex}^{\rm ref}=$ 30 mT. The angle dependence displays the well-known butterfly shape, which has been reported in the SAW-based magnetostrictive systems \cite{weiler_adfmr,dreher_adfmr,labanowski_adfmr}. It should be noted that the SAW amplitude is minimized at $\mu_{0} H_{\rm ex}$ = 2 mT and $\phi_{\rm h}$ = 45$^{\circ}$ and the absorption level is much larger than in previous reports. This is ascribed to the resonance frequency of the magnetic precession ($\omega_{\rm m}$) approaching $\omega_{\rm a}$ so that efficient spin-wave driving using the acoustic vibrations becomes possible in the cavity magnomechanical system [See also the simulated result shown in Supplemental Information (section II)].\\
\hspace*{0.5cm}Before showing the detailed field response, we theoretically confirm the static magnetization dynamics of the Ni film for three specific field directions ($\phi_{\rm h}$ = 90$^\circ$, 45$^\circ$, and 0$^\circ$). The magnetization dynamics of the Ni film is mainly governed by the external magnetic field ($\mu_{0} \mbox{\boldmath $H_{\rm ex}$}$), in-plane uniaxial anisotropy ($\mbox{\boldmath $B_{\rm u}$}$), and out-of-plane shape anisotropy ($\mbox{\boldmath $B_{\rm d}$}$). Figure 2(a) displays the relation of these vectors in an xy-coordinate system, where $\phi_{\rm h}$, $\phi_{\rm m}$, and $\phi_{\rm u}$ are the angle of $\mu_{0} \mbox{\boldmath $H_{\rm ex}$}$, magnetization vector $\mbox{\boldmath $M$}$, and the easy axis of in-plane uniaxial anisotropy, respectively, from the SAW propagation direction ($x$-axis). The magnetic free energy density normalized to the saturation magnetization ($M_{\rm s}$) is given by
\begin{equation}
G=-\mu_{0} \mbox{\boldmath $H_{\rm ex} \cdot m$} -B_{\rm u} \mbox{\boldmath ($m \cdot u$)}^{2}+B_{\rm d}m_z^{2},
\end{equation}
where, $\mbox{\boldmath $m$}$ and $\mbox{\boldmath $u$}$ are the unit vectors of magnetization ($\mbox{\boldmath $M$}$) and easy axis of in-plane uniaxial anisotropy. The equilibrium position of the magnetization is determined so as to minimize $G$.\\
\hspace*{0.5cm}The magnetization component projected to the field at $\phi_{\rm h}$ = 90$^\circ$, 45$^\circ$, and 0$^\circ$ are simulated by assuming $\phi_{\rm u}$ = 25$^\circ$ as shown in Fig. 2(c). When the field is strong enough, e.g., $\mu_{0} H_{\rm ex}$ = 30 mT, the magnetization is perfectly aligned to it. As the field decreases in the backward sweep, the magnetization is oriented toward the uniaxial anisotropy axis from the field direction. At $\phi_{\rm h}$ = 0$^\circ$ (45$^\circ$), the rotation starts from $\phi_{\rm m}$ = 0$^\circ$ (45$^\circ$) around 2 mT, and $\phi_{\rm m}$ is aligned to $\phi_{\rm u}$ = 25$^\circ$ at 0 mT. The rotation continues to $\phi_{\rm m}$ = 45$^\circ$ (0$^\circ$), and then it is inverted and finally re-oriented to the negative field, i.e. $\phi_{\rm m}$ = 180$^\circ$ (225$^\circ$). On the other hand, this rotation gradually starts from 10 mT at $\phi_{\rm h}$ = 90$^\circ$, in which the magnetization angle gradually changes from $\phi_{\rm m}$ = 90$^\circ$ to 10$^\circ$ passing through $\phi_{\rm u}$ = 25$^\circ$ and inversion occurs at -0.6 mT. Even after the inversion, the rotation continues from 215$^\circ$ to the field direction i.e. 270$^\circ$. This simulation reveals that the magnetization is not always parallel to the field and its orientation is governed by the relationship between the external and anisotropic fields. In analyzing the magnetomechancial coupling effect, the effect of this magnetization rotation should be taken into account. This static magnetization dynamics is described in more detail in Supplemental Information (Section III).\\
\hspace*{0.5cm}We then measured the field response of $S_{21}$ for these three field directions as shown in Fig. 3(a)-(c). We also compared the response of six different FP resonances, P1--P6, to confirm the effect of the quality factor, i.e., the cavity confinement. For all three magnetic field directions, the backward field sweep from $\mu_0H_{\rm ex}$ = 30 mT reduces the acoustic magnitude because of an enhanced efficiency of the magnetostrictive spin driving. It is remarkable that the acoustic absorption power is amplified as $Q_{\rm a}$ increases and exceeds $80\%$ at these angle configurations, which is much larger than in previous systems ($< 10\%$) \cite{weiler_adfmr,dreher_adfmr}. Obviously, the enhancement of the acoustic absorption is a consequence of the cavity effect of our magnomechanical system. The results are also discussed in Supplemental Information (Sections IV and V).\\
\hspace*{0.5cm}The different field responses observed for three field directions are explained by the magnetization dynamics already discussed [Fig. 2(c)]. We need to take into account two contributions. One is the variation of magnetostriction coupling constant $J$ defined by
\begin{equation}
J = k b_{\rm ma} \sin2\phi_{\rm m} V_{\rm ma},
\end{equation}
Here, $b_{\rm ma}$ is the magnetoelastic coupling constant, and $V_{\rm ma}$ is the effective mode overlap of the magnetostriction. Obviously, the absolute value of $J$ changes with magnetization rotation and becomes maximum at $\phi_{\rm m}$ = 45$^\circ(2n+1)$ ($n$ = 0,1,2...). The other contribution is the spin resonance effect, where the maximum efficiency is obtained when magnonic resonance frequency $\omega_{\rm m}$ matches the acoustic drive frequency $\omega$. The displacement amplitude ($u_{\rm a}$) of acoustic waves is given by,
\begin{equation}
u_{\rm a}(\omega) = \frac{F_{\rm d}\rho_{\rm a}^{-1}}{-(\omega^{2}-\omega_{\rm a}^{2})-i\omega\kappa_{\rm a}-J^{2}\chi_{\rm m}},
\end{equation}
where $\chi_{\rm m}=\frac{2M_{\rm s}\gamma\eta\omega_{\rm m}}{\rho_{\rm a}V_{\rm a}V_{\rm m}(1+\alpha^2)(-\omega^{2}+\omega_{\rm m}^{2}+\frac{\kappa_{\rm m}^{2}}{4}-i\kappa_{\rm m}\omega)}$ is the magnonic susceptibility. In the expressions, $F_{\rm d}$, $\rho_{\rm a}$, and $\gamma$ are the driving force density, mass density, and gyromagnetic ratio, and $V_{\rm a}$ ($V_{\rm m}$) and $\eta$ denote the effective mode volume of the acoustic (magnonic) system and the magnetoelastic coefficient, respectively. The resonance angular frequency and damping rate of the acoustic (magnonic) system are defined by $\omega_{\rm a}$ and $\kappa_{\rm a}$ ($\omega_{\rm m}$ and $\kappa_{\rm m}$), respectively. The detailed derivation of Eq. (3) and the parameters are found in Methods. The field response of SAW absorption efficiency is determined by taking into account these two contributions: the couping enhancement by spin resonance condition and the change in $J$ caused by the field-induced magnetization rotation.\\
\hspace*{0.5cm}From Eqs. (2) and (3), the dynamics of the cavity magnomechanical system illustrated in the inset of Fig. 1(a) was simulated using the fitting parameters as described in Methods. The right panel of Fig. 3(a)-(c) shows the calculation results, where the normalized amplitude is plotted on the vertical axis. The calculation well reproduces the experimental results on the field and quality-factor dependencies for all three field directions even though they showed different magnetization dynamics. This analysis indicates that the high magnomehcanical coupling is obtained by applying the spin resonance condition while maintaining $\phi_{\rm m}\sim$ 45$^\circ$ to obtain large magnetostriction effects.\\
\hspace*{0.5cm}The strong driving of the ferromagnetic spin-wave oscillations enables the acoustic dynamics to be modulated via back-action. The spectral response in P1 resonance is measured with backward field sweep at $\phi_{\rm h}$ = 90$^\circ$, 45$^\circ$, and 0$^\circ$ as shown in the left panel of Fig. 4(a), (c) and (e), and experimental quality-factor $Q_{\rm a}^{\prime}$ and resonance frequency $\omega_{\rm a}^{\prime}/(2\pi)$ acquired from the measured spectrum are plotted in Fig. 4(b), (d) and (f) respectively. At $\phi_{\rm h}$ = 45$^\circ$, $\omega_{\rm a}^{\prime}$ and $Q_{\rm a}^{\prime}$ are modulated with decreasing field and changes in polarity at $\mu_{0} H_{\rm ex} = \pm$2.1 mT. Similar field dependencies are also observed at $\phi_{\rm h}$ = 90$^\circ$, but the width of the dips becomes narrow compared to that at $\phi_{\rm h}$ = 45$^\circ$. The spin-wave oscillation is generated only in the magnetization rotation regime $|\mu_{0} H_{\rm ex}| < 10$ mT in $\phi_{\rm h}$ = 90$^{\circ}$, allowing the magnetostriction to be activated only by those fields ($J\neq0$), and then $J$ becomes a maximum when $\phi_{\rm m}$ = 45$^{\circ}$ and 225$^{\circ}$ at the fields of the dips. On the other hand, this nonzero $J$ is kept in almost the entire range between $\pm$30 mT at $\phi_{\rm h}$ = 45$^{\circ}$ because initially $\phi_{\rm m}=\phi_{\rm h}$. Hence, the available field for the back-action is limited in Figs. 4(b). Moreover, a different feature of the field response is observed at $\phi_{\rm h}$ = 0$^{\circ}$ in Fig. 4(f), where a single dip appears at -1 mT. This is because $\phi_{\rm m}$ = 45$^{\circ}$ only around the field. Since these experimental behaviors are reproduced by the theoretical model, the variation in $\omega_{\rm a}^{\prime}$ and $Q_{\rm a}^{\prime}$ result from the dynamic back-action of the spin-wave excitation, which more than doubles the acoustic damping rate.\\
\begin{figure}[t]
	\begin{center}
		\vspace{-0.0cm}\hspace{-0.0cm}
		\includegraphics[scale=1]{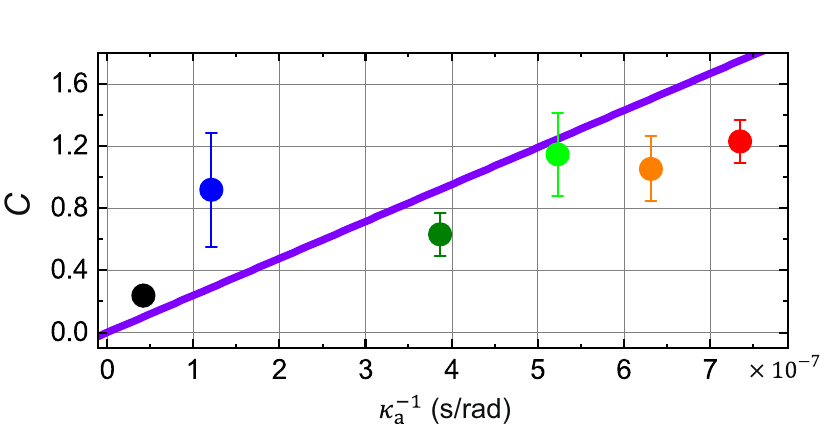}
		\vspace{-0cm}
		\caption{Inverse of acoustic damping $\kappa_{\rm a}^{-1}$ dependence of the cooperativity $C$, experimentally obtained by measuring the change in total acoustic damping $\kappa_{\rm a}^{\prime}$ between $\mu_{0} H_{\rm ex}^{\rm res}$ = 0.6 mT and 1.8 mT at $\phi_{\rm h}$ = 90$^\circ$, as denoted by solid circles whose colors correspond to those of P1--P6 resonances [see Fig. 1(d)]. The theoretical prediction of $C$ obtained from the Eqs. (4) and (5) is plotted by a purple solid line.}
		\label{fig 1}
		\vspace{-0.5cm}
	\end{center}
\end{figure}
\hspace*{0.5cm}The cavity effect on the magnetostrictive coupling is quantitatively investigated as function of the acoustic damping. To do so, we use cooperativity parameter $C$, the ratio of coherent magnon-phonon coupling rate ($g$) to $\kappa_{\rm a}$ and $\kappa_{\rm m}$, which expresses the efficiency of coherent energy transfer between different physical systems and is commonly used in cavity optomechanics and c-QAD. In our cavity magnomechanical system, 
\begin{equation}
C = \frac{\kappa_{\rm a}^{\rm \prime}}{\kappa_{\rm a}}-1 = \frac{4g^{2}}{\kappa_{\rm m} \kappa_{\rm a}},
\end{equation}
where magnon-phonon coupling $g$ is
\begin{equation}
g=J\sqrt{\frac{M_{\rm s}\gamma \eta}{2V_{\rm a}V_{\rm m}\rho_{\rm a}(\omega_{\rm a}
		+\frac{\kappa_{\rm m}^{2}}{16\omega_{\rm a}})(1+\alpha^2)}},
\end{equation}
with the total damping rate $\kappa_{\rm a}^{\rm \prime}$. To obain Eq. (4), we adopt the approximation that variation in $\omega_{\rm a}^{\prime}$ is negligibly smaller than that in $\kappa_{\rm a}^{\prime}$. The expression of $g$ including $\eta$ is derived from Eq. (3) only when the condition $\omega=\omega_{\rm a}=\omega_{\rm m}$ is satisfied (the detail explanation can be found in Methods). Note that our system at $\phi_{\rm h}=90^{\circ}$ can simultaneously fulfill the resonance coupling condition $\omega_{\rm m} \approx \omega_{\rm a}$ and the maximum magnetostriction $\phi_{\rm m} \approx 45^{\circ}$ in $\mu_{0} H_{\rm ex}$ = 0.6 $\sim$ 1.8 mT (see Supplemental Information Section VI). By extracting $\kappa_{\rm a}^{\rm \prime}$ from P1--P6 resonances at the field angle, $C$ is plotted as function of $\kappa_{\rm a}^{-1}$ as shown in Fig. 5. The $C$ calculated from Eqs. (4) and (5) is also shown by a solid line. The experimental $C$ increases with increasing $\kappa_{\rm a}^{-1}$, namely $Q_{\rm a}$, which is consistent with the calculation result. Finally, the cooperativity  reaches $C$ = 1.2$\pm$0.1 in P1 resonance with $Q_{\rm a}$ = 4,500, thus indicating coherent interaction between spin and acoustic waves. The coupling rate in our system is estimated to be $g = 2\pi \times$ (9.9$\pm$0.2) MHz, which is larger than the intrinsic acoustic damping but smaller than the magnetic one ($\kappa_{\rm m} > g > \kappa_{\rm a}$). Employing a low-magnetic damping material such as yttrium-iron-garnet (YIG) as a ferromagnet will be valid way to further increase the effective interaction and achieve strong coupling condition ($g > \kappa_{\rm a}, \kappa_{\rm m}$).\\
\hspace*{0.5cm}With cavity magnomechanics, we have an opportunity to further improve functionality. For instance, combining it with spintronics technology, where ferromagnetic dynamics can be engineered by advanced microfabrication, will enable us to design the magnetoelastic effect, which will lead to robust and versatile spin excitation schemes unrestricted by the field configuration. The integration of a wavelength-scale phononic-crystal structure in the system will be useful for enhancing the ability to spatially control acoustic waves \cite{safavi_pnc2,hatanaka_hyperPnC}, allowing the development of large-scale magnomechanical circuits. Thus, our cavity magnomechanical system will open up the possibility of controlling acoustic phonons with magnons and vice versa, which holds promise for establishing novel magnon-phonon technologies for classical and quantum signal processing applications.\\ 
%

%
\newpage
\section{Methods}
\subsection{Fabrication and measurement}
\hspace*{0.0cm}The device was fabricated on a commercially available 128$^\circ$-$Y$ LiNbO$_{3}$ substrate (Black-LN, Yamaju Ceramics Co., Ltd.). A rectangular Ni thin film with a thickness of $d$ = 50 nm was formed. Nickel was chosen because is has the highest magnetostrictive constant among common ferromagnets such as cobalt, permalloy and iron. In the simulation, we considered out-of-plane shape anisotropy ($B_{\rm d}$) and in-plane uni-axial anisotropy ($B_{\rm u}$) in the ferromagnet. The former is caused by the thin-film structure, and the latter can occur during the film growth on the substrate accidentally or could be caused by the rectangular shape of the film. To sandwich it, two IDT electrodes with a thickness of 50 nm were introduced. They have 19 finger pairs with a pitch of $2p$ = 4.0 $\mu$m. They were used to excite and detect SAW transmission piezoelectrically. Acoustic Bragg reflectors were built to confine the acoustic waves. They consisted of 568-periodicity strip lines with a pitch of $p$ = 2.0 $\mu$m and were connected to electrical ground. Acoustic trapping in this cavity is a result of constructive interference between incident waves and reflected waves from the periodic metallic strip lines. These Bragg reflectors and IDTs are made of aluminum (Al) because the low mass density among metals allows gentle confinement of SAWs due to the relatively small acoustic reflectivity and suppresses coupling of waves reflected to bulk acoustic modes.\\
\hspace*{0.5cm}Resonant SAWs were measured from an IDT electrode in $S_{21}$ with a network analyzer (E5080A, Keysight) by injecting microwave signals with -20 dBm into the other one. Spurious electromagnetic waves due to crosstalk between the IDTs were filtered out by a time-gating technique. Simultaneously, a static magnetic field was applied in the plane of Ni thin film from an electromagnetic coil, and the field was swept backward, from $\mu_{0} H_{\rm ex}$ = 30 mT to 0 mT or -30 mT, in the experiments in Figs. 2, 3 and 4. All the measurements in this study were performed at room temperature and in a moderate vacuum ($\sim 10^{1}$ Pa).
\subsection{Description of the dynamics via magnon-phonon coupling}
Dynamics of the acoustic mode interacting with the magnetization as macrospins via the magnetostrictive coupling is given by the following equation of motion,
\begin{align}
\begin{split}
\rho_{\rm a}\ddot{U}_{\rm a}(t)&+\rho_{\rm a}\kappa_{\rm a}\dot{U}_{\rm a}(t)+\rho_{\rm a}\omega_{\rm a}^2U_{\rm a}(t)\\
&-M_{\rm s}b_\mathrm{ma}\sin 2\phi_{\rm m}\frac{ V_\mathrm{a-m_2}}{V_{\rm a}}m_2(t)=F_{\rm d}
\end{split}
\end{align}
where $U_{\rm a}$, $\kappa_{\rm a}$, and $\omega_{\rm a}$ are the displacement, damping rate, and angular frequency of the acoustic mode, and $\rho_{\rm a}$ is the density of the ferromagnet. The magnetostrictive coupling between the acoustic mode $U_{\rm a}(t)$ and the magnetic precession component $m_2(t)$ (see Fig. 2(a)) is described in the fourth term on the left-hand side with the saturation magnetization $M_{\rm s}$, magnetoelastic coupling constant $b_\mathrm{ma}$, magnetization angle $\phi_{\rm m}$, mode volume of the acoustic mode $V_{\rm a}\equiv \int\mathrm{d}V \rho_{\rm a}({\bf r})[\Psi_{\rm a}({\bf r})]^2$, and the overlap integral $V_\mathrm{a-m_2}\equiv \int\mathrm{d}V\left[\Psi_{\rm a}({\bf r})\partial_x \Psi_{\rm m_2}({\bf r})\right]$, with the spatial distribution of acoustic mode $\Psi_{\rm a}({\bf r})$ and that of macrospin $\Psi_{\rm m_2}({\bf r})$. Note that each spatial distribution is normalized so that $\mathrm{max}_{\bf r}\left[\Psi_j({\bf r})\right]=1$. The driving force from the IDT electrodes is simply denoted by $F_{\rm d}$.\\
On the other hand, the dynamics of macrospins $m_1(t)$ and $m_2(t)$ are given by the linearized Landau-Lifshitz-Gilbert (LLG) equations \cite{weiler_adfmr,dreher_adfmr},
\begin{align}
\frac{\alpha}{\gamma} \dot{m}_1(t)+(G_{11}-G_3) m_1(t)-\frac{1}{\gamma} \dot{m}_2(t)=0,
\end{align}
\begin{align}
\begin{split}
\frac{\alpha}{\gamma} \dot{m}_2(t)&+(G_{22}-G_3)m_2(t)+\frac{1}{\gamma} \dot{m}_1(t)\\
&+b_\mathrm{ma}\sin 2\phi_{\rm m} \frac{V_\mathrm{m_2-a}}{V_{\rm M}}U_{\rm m}(t)=0.
\end{split}
\end{align}
where $\alpha$ and $\gamma$ are the Gilbert damping factor and gyromagnetic ratio of the Ni film, respectively. Each coefficient is given by $G_{11}=2B_{\rm d}$, $G_{22}=-2B_{\rm u} \sin^2(\phi_{\rm m}-\phi_{\rm u})$, and $G_3=-\mu_0 H_\mathrm{ex}\cos(\phi_{\rm m}-\phi_{\rm h})-2B_{\rm u} \cos^2(\phi_{\rm m}-\phi_{\rm u})$ with the in-plane uniaxial anisotropy $B_{\rm u}$, and out-of-plane shape anisotropy $B_{\rm d}$, along with the angle of the external magnetic field $\phi_{\rm h}$, that of the magnetization $\phi_{\rm m}$, and that of the uniaxial anistropic field $\phi_{\rm u}$ [see Fig. 2(a)]. The magnetostrictive coupling is given by the fourth term including the overlap integral $V_\mathrm{m_2-a}=\int\mathrm{d}V[\Psi_{\rm m}({\bf r})\partial_x \Psi_{\rm a}({\bf r})]$, and the mode volume of the macrospin, $V_{\rm m}\equiv \int \mathrm{d}V [\Psi_{\rm m_2}({\bf r})]^2 = \int \mathrm{d}V [\Psi_{\rm m_1}({\bf r})]^2$. Here we assume that the spatial distribution of the macrospins are completely determined by that of strain, i.e., $\Psi_{\rm m_1}({\bf r})\approx \Psi_{\rm m_2}({\bf r})\approx \partial_x \Psi_{\rm a}({\bf r})$. The LLG equations can be diagonalized to
\begin{equation}
\begin{split}
\dot{m}_\pm=&\left(\pm i\omega_{\rm m}-\frac{\kappa_{\rm m}}{2}\right)m_\pm\\
 &\pm b_\mathrm{ma}\sin 2\phi_{\rm m} \frac{V_\mathrm{m_2-a}}{V_{\rm m}}\frac{\gamma }{1+\alpha^2}\eta U_{\rm a},
\end{split}
\end{equation}
where
\begin{align}
\omega_{\rm m}=& \frac{\sqrt{4\tilde{G}_1\tilde{G}_2-\alpha^2(\tilde{G}_1-\tilde{G}_2)^2}}{2},\\
\kappa_{\rm m}=&\alpha(\tilde{G}_1+\tilde{G}_2)
\end{align}
and
\begin{align}
\eta=&-i\frac{\alpha}{2}-\frac{2\tilde{G}_1+(\tilde{G}_1-\tilde{G}_2)\alpha^2}{2\sqrt{4\tilde{G}_1\tilde{G}_2-\alpha^2(\tilde{G}_1-\tilde{G}_2)^2}}
\end{align}
are defined with $\tilde{G}_{j}\equiv \frac{\gamma}{1+\alpha^2}\left(G_{jj}-G_3\right)$. Because $m_2=i(m_+ +m_-)$, the equation of motion for the acoustic mode can be rewritten as
\begin{align}
\begin{split}
&\rho_{\rm a}\ddot{U}_{\rm a}(t)+\rho_{\rm a}\kappa_{\rm a}\dot{U}_{\rm a}(t)+\rho_{\rm a}\omega_{\rm a}^2U_{\rm a}(t)\\
&-iM_{\rm s}b_\mathrm{ma}\sin 2\phi_{\rm m}\frac{ V_\mathrm{a-m_2}}{V_{\rm a}}\left[m_+(t)+m_-(t)\right]=F_{\rm d}.
\end{split}
\end{align}
Because the acoustic mode is spatially distributed as a standing wave on the Ni film, i.e., $\Psi_{\rm a}({\bf r})=\tilde{\Psi}_{\rm a}(y,z)\sin k x$, we can simplify the spatial overlap as follows:
\begin{align}
\begin{split}
V_\mathrm{a-m_2}=&-k\int\mathrm{d}y\mathrm{d}z \int_{-w/2}^{w/2}\mathrm{d}x\sin^2k x=-\frac{kw}{2}\int\mathrm{d}y\mathrm{d}z\\
=&-kV_\mathrm{ma}
\end{split}
\end{align}
\begin{align}
\begin{split}
V_\mathrm{m_2-a}=&k\int\mathrm{d}y\mathrm{d}z \int_{-w/2}^{w/2}\mathrm{d}x
\cos^2k x=\frac{kw}{2}\int\mathrm{d}y\mathrm{d}z \\
=&kV_\mathrm{ma}
\end{split}
\end{align}
where $k$ is the wavevector of the acoustic mode and $w$ is the width of Ni film along the propagation direction with the structural condition $w\gg k^{-1}$. By denoting $J\equiv kb_\mathrm{ma}\sin 2\phi_{\rm m} V_\mathrm{ma}$, we achieve the simplified expression,
\begin{align}
\begin{split}
\ddot{U}_{\rm a}(t)&+\kappa_{\rm a}\dot{U}_{\rm a}(t)+\omega_{\rm a}^2U_{\rm a}(t)\\
&+i\frac{M_{\rm s}J}{\rho_{\rm a}V_{\rm a}}\left[m_+(t)+m_-(t)\right]=F_{\rm d}.\label{phonon_t1}
\end{split}
\end{align}
\begin{align}
&\dot{m}_\pm=\left(\pm i\omega_{\rm m}-\frac{\kappa_{\rm m}}{2}\right)m_\pm \pm  \frac{J}{V_{\rm m}}\frac{\gamma }{1+\alpha^2}\eta U_{\rm a}.\label{magnon_t1}
\end{align}
Thus, by performing Fourier transform on Eqs. (\ref{phonon_t1}) and (\ref{magnon_t1}), the displacement amplitude of acoustic waves ($u_{\rm a}$) is given by
\begin{equation}
u_{\rm a}(\omega) = \frac{F_{\rm d}\rho_{\rm a}^{-1}}{-(\omega^{2}-\omega_{\rm a}^{2})-i\omega\kappa_{\rm a}-J^{2}\chi_{\rm m}},
\end{equation}
with the magnonic susceptibility
\begin{equation}
\chi_{\rm m}=\frac{2M_{\rm s}\gamma\eta\omega_{\rm m}}{\rho_{\rm a}V_{\rm a}V_{\rm m}(1+\alpha^2)(-\omega^{2}+\omega_{\rm m}^{2}+\frac{\kappa_{\rm m}^{2}}{4}-i\kappa_{\rm m}\omega)}.
\end{equation}\\
The representation of phonon-magnon coupling constant $g$ is unveiled by performing the rotating frame approximation with transformations of $U_{\rm a}=\sqrt{\frac{\hbar}{2\omega_{\rm a}\rho_{\rm a} V_{\rm a}}}A(t)e^{-i\omega_{\rm a} t}+\mathrm{c.c.}$ and $m_\pm=\sqrt{\frac{\hbar \gamma \eta}{M_{\rm s}V_{\rm m}(1+\alpha^2)}}M_\pm(t) e^{\pm i\omega_{\rm m}t}$ as follows:
\begin{align}
\dot{A}(t)=&-\frac{\kappa_{\rm a}}{2}A(t)+gM_-(t)+\tilde{F}_{\rm d}(t)\\
\dot{M}_-(t)=&-\frac{\kappa_{\rm m}}{2}A(t)-g A(t),
\end{align}
with
\begin{align}
g=&J\sqrt{\frac{M_{\rm s}\gamma \eta}{2V_{\rm a}V_{\rm m}\rho_{\rm a}(\omega_{\rm a}+\frac{\kappa_{\rm m}^{2}}{16\omega_{\rm a}})(1+\alpha^2)}}.
\end{align} 
In case of $\omega_{\rm a} \gg \kappa_{\rm m}$, it can be simplified to
\begin{align}
g \approx &J\sqrt{\frac{M_{\rm s}\gamma \eta}{2V_{\rm a}V_{\rm m}\omega_{\rm a}\rho_{\rm a}(1+\alpha^2)}}\\
\approx & kb_{\rm ma}\sin{2\phi_{\rm m}}\sqrt{\frac{M_{\rm s}\gamma \eta}{2\omega_{\rm a}\rho_{\rm a}(1+\alpha^2)}\frac{w}{L_\mathrm{c}}\frac{d}{\lambda}} \label{g_rep}
\end{align}
where $\tilde{F}_{\rm d}$ is the resonance driving force in the rotating frame, $L_\mathrm{c}$ is the acoustic cavity length along the propagation direction, $d$ is the Ni film thickness, and $\lambda$ is the wavelength of the acoustic mode approximated to be the penetration depth of the acoustic mode in the Ni film. Note that the final approximation in Eq. (\ref{g_rep}) is obtained by assuming that $\Psi_{\rm a}({\bf r})\approx D_0e^{-z/\lambda}\sin kx $ with an arbitrary constant $D_0$. Here, we emphasize that $A(t)$ and $M(t)$ have no dimension unit, which implies that the quantization of each amplitude brings them to the description for quantum dynamics as well as the formulation given in magnon-phonon hybrid quantum systems \cite{agarwal_magnomech,li_njp,li_PRRsearch}.
\subsection{Acoustic and magnetic parameters for simulations}
All the parameters used in the simulations are shown in the table below, and they are in good agreement with previous reports \cite{weiler_adfmr,dreher_adfmr}. Only the value of $\phi_{\rm u}$ is chosen in such a way that this gives a best fitting of Eq. (3) to the experimental data.
\begin{table}[h]
	\begin{tabular}{|c|c|c|} \hline
		$b_{\rm ma}$ & Magnetoelastic coupling constant & 14 T \\
		$B_{\rm d}$ & Out-of-plane shape anisotropy & 0.21 T \\
		$B_{\rm u}$ & In-plane uniaxial anisotropy & 1.8 mT \\
		$\phi_{\rm u}$ & Angle of in-plane uniaxial anisotropy & 25$^{\circ}$ \\
		$\alpha$ & Gilbert damping factor & 0.08 \\
		$M_{\rm s}$ & Saturation magnetization & 370 kA/m \\
		$\gamma$ & Gyromagnetic ratio & 2.185$\hbar/\mu_{\rm B}$ \\
		$\rho_{\rm a}$ & Mass density & 8900 kg/m$^{3}$ \\ \hline
	\end{tabular}
\end{table}
\section{Acknowledgments}
The authors thank H. Murofushi and S. Sasaki for support in the nickel deposition and sample preparation.
\section{Author contributions}
D.H. fabricated the device and performed the measurements and the data analysis. M.A. made the theoretical model, and D.H. M.A and H.Y. conducted the simulations with support from H.O.. D.H., H.Y. and M.A. wrote the manuscript. All authors discussed the results through paper preparation.

\end{document}